\documentclass[10pt,twocolum,final]{IEEEtran}
\usepackage[utf8]{inputenc}
\usepackage{bm}
\usepackage{amsmath}
\usepackage{amssymb}
\usepackage{graphicx}

\usepackage[
            linkcolor=black, 
            anchorcolor=black, 
            citecolor=black, 
            urlcolor=black
           ]{hyperref}

\makeatletter

\usepackage{amsfonts}
\usepackage{color}
\usepackage{multicol}
\usepackage{subfigure}
\usepackage{bm}
\usepackage{latexsym}
\usepackage{stfloats}
\usepackage{float}
\usepackage{exscale}
\usepackage{relsize}
\usepackage{cite}
\usepackage{cases}
\usepackage{algorithm}
\usepackage{algpseudocode}
\usepackage{graphics}
\usepackage{epsfig}
\usepackage{epstopdf}
\usepackage{breqn}
\usepackage{enumerate}
\usepackage{multirow}
\usepackage{caption}
\usepackage{xcolor}
 \usepackage{array}

\makeatother

\vspace{0.5cm}
\begin{document}
\title{Robust Transmission Design for RIS-Aided Communications with Both Transceiver Hardware Impairments and Imperfect CSI\vspace{-0.2em}}
\IEEEoverridecommandlockouts
\author{
        Zhangjie~Peng,
        Zhiwei Chen,
        Cunhua Pan,~\IEEEmembership{Member,~IEEE},
        Gui Zhou,
        and Hong Ren,~\IEEEmembership{Member,~IEEE}
\vspace{-0.2em}
\vspace{-0.5cm}

 \thanks{\emph{ (Corresponding authors: Cunhua Pan; Zhiwei Chen.)}}
\thanks{Zhangjie Peng is with the College of Information, Mechanical, and Electrical Engineering, Shanghai Normal University, Shanghai 200234, China, also with the National Mobile Communications Research Laboratory, Southeast University, Nanjing 210096, China, and also with the Shanghai Engineering Research Center of Intelligent Education and Bigdata, Shanghai Normal University, Shanghai 200234, China (e-mail: pengzhangjie@shnu.edu.cn).}
 \thanks{Zhiwei Chen is with the College of Information, Mechanical and Electrical Engineering,
 Shanghai Normal University, Shanghai 200234, China (e-mail: 1000497437@smail.shnu.edu.cn).}
 \thanks{Cunhua Pan is with the National Mobile Communications Research Laboratory, Southeast University, Nanjing 210096, China. He was with the School of
Electronic Engineering and Computer Science at Queen Mary University of London, London E1 4NS, U.K. (e-mail: cunhuapan21@gmail.com).}
 \thanks{Gui Zhou is with the School of Electronic Engineering and Computer Science, Queen Mary University of London, London  E1  4NS, U.K.  (e-mail: g.zhou@qmul.ac.uk). }
 \thanks{Hong Ren is with the National Mobile Communications Research Laboratory, Southeast University, Nanjing 210096, China (e-mail: hren@seu.edu.cn).
}
}

\maketitle

\newtheorem{lemma}{Lemma}
\newtheorem{theorem}{Theorem}
\newtheorem{remark}{Remark}
\newtheorem{corollary}{Corollary}
\newtheorem{proposition}{Proposition}
\newcounter{TempEqCnt}
\vspace{-0.5cm}
\begin{abstract}
Reconfigurable intelligent surface (RIS) or intelligent reflecting surface (IRS)
has recently been envisioned as one of the most promising technologies in the future sixth-generation (6G) communications.
In this paper, we consider the joint optimization of the transmit beamforming at
the base station (BS) and the phase shifts at the RIS for
an RIS-aided wireless communication system with both hardware impairments and imperfect channel state information (CSI).
 Specifically,
 we assume both the BS-user channel and the BS-RIS-user channel are imperfect due to the channel
estimation error,
and we consider the channel estimation error under the statistical CSI error model.
Then, the transmit power of the BS is minimized, subject to the outage probability constraint and the unit-modulus constraints on the reflecting elements.
By using Bernstein-type inequality and semidefinite relaxation
(SDR) to reformulate the constraints, we transform the
optimization problem into a semidefinite programming (SDP) problem.
Numerical results show that the proposed robust design algorithm  can ensure  communication quality of the user in the presence of both hardware impairments and imperfect CSI.


\end{abstract}

\section{Introduction}

Due to the rapid development of metamaterials,
reconfigurable intelligent surface (RIS),
which is composed of multiple reflecting units,
has recently emerged as a promising technique to improve the quality of the future wireless communications\cite{9318531,9366346,8811733,9558795}.
Specifically,
by controlling the reflecting elements at the RIS, 
the reflection direction of the electromagnetic wave can be controlled accurately.
 Then, the reflected signals can be reconfigured to propagate towards their desired directions.
As a result, RIS has been proposed to be employed in a variety of communication scenarios such as
 mobile edge computing\cite{9133107}, secrecy communication\cite{9201173}, and unmanned aerial vehicle-assisted communications\cite{9367288}.

However, most of the existing works on RIS-aided communications are based on the assumption of perfect transceiver hardware and perfect channel state information (CSI),
which may not be realistic in practice.
Performance degradation will be incurred  by inevitable hardware impairments
such as hardware aging, imperfect power amplifier, oscillators noise, low-resolution digital-to-analog converters (DACs) and imperfect analog-to-digital converters (ADCs)\cite{9239335,9159653}.
It is noted that hardware impairments will
impair the signal quality at the receiver.
In\cite{9239335},
a robust beamforming  scheme was proposed
for an RIS-aided point-to-point wireless system with hardware impairments.
The authors in \cite{9159653} demonstrated that the hardware impairments limited the performance of
 RIS-aided systems and  highlighted  the importance of accurately modeling the transceiver hardware impairments.
However, the above-mentioned works \cite{9239335,9159653} were based on the assumption of  perfect CSI.
Due to the passive property of the RIS,
it is challenging to acquire accurate CSI.
Recently, some existing works have studied the impact of
imperfect CSI on the performance of the RIS-aided wireless systems\cite{9180053,9534477}.
The authors in\cite{9180053} proposed a robust beamforming design scheme for
an RIS-aided MISO system, where
the imperfect cascaded CSI was taken into account.
In\cite{9534477},
the authors investigated the energy efficiency
and power scaling laws of RIS-aided systems with imperfect CSI and transceiver hardware impairments.
However, the existing literature  only considered imperfect CSI or only imperfect hardware.
There are no literature studying the similar optimization problems considering both the imperfect CSI and the transceiver hardware impairments.

\raggedbottom
Against the above background, 
we study the robust transmission scheme that takes into account
both  transceiver hardware impairments and imperfect CSI.
The main contributions of this work are summarized as follows: $1)$
To be specific,
by jointly optimizing the transmit beamforming vector at the base station (BS) and the phase shifts at the RIS,
the transmit power of the BS is minimized,
subject to the outage probability constraint and the unit-modulus
constraints on the reflecting elements.
$2)$ By using Bernstein-Type inequality, we effectively simplify the rate outage probability constraint. Then, the beamforming vectors  are obtained by utilizing the
semidefinite relaxation (SDR).
$3)$ Simulation results show that the  proposed robust design
can guarantee the quality of service requirements
in an RIS-aided MISO wireless system with both  transceiver hardware impairments and imperfect CSI.

{\setcounter{equation}{3} \label{20}}
\begin{figure*}[hb]
\hrulefill
\begin{align}
 \mathbb{E}\{n_{\mathrm e} n_{\mathrm e}^*\} \nonumber
 =&    \mathbb{E}\{(({\bf{g}}^{\mathrm H}+{\bf e}^{\mathrm H}\mathbf{Q}){\bf n}_{\mathrm t}+n+n_{\mathrm r})({\bf n}_{\mathrm t}^{\mathrm H}(\mathbf{Q}^{\mathrm H}{\bf e}+{\bf{g}})+n^{*}+n_{\mathrm r}^{*})\} \nonumber\\
 = & ({\bf{g}}^{\mathrm H}+ {\bf e}^{\mathrm H}\mathbf{Q})\mathbb{E}\{{\bf n}_{\mathrm t}{\bf n}_{\mathrm t}^{\mathrm H}\}(\mathbf{Q}^{\mathrm H}{\bf e}+{\bf{g}})\nonumber\\
 & \!+\!\mathbb{E}\{nn^{*}\}+\!\beta_{r}\mathbb{E}\{(({\bf{g}}^{\mathrm H}\!+\!{\bf e}^{\mathrm H}\mathbf{Q}){\bf v}s\!+\!({\bf{g}}^{\mathrm H}+{\bf e}^{\mathrm H}\mathbf{Q}){\bf n}_{\mathrm t}\!+\!n)(s^{*}{\bf {\bf v}}^{\mathrm H}(\mathbf{Q}^{\mathrm H}{\bf e}\!+\!{\bf{g}})\!+\!{\bf n}_{\mathrm t}^{\mathrm H}(\mathbf{Q}^{\mathrm H}{\bf e}\!+\!{\bf{g}})\!+n^{*})\}\nonumber\\
 = & ({\bf{g}}^{\mathrm H}+ {\bf e}^{\mathrm H}\mathbf{Q})\left(\beta_{ r}{\bf v}{\bf {\bf v}}^{\mathrm H}+(1+\beta_{ r})\beta_{ t}{\textrm{diag}}({\bf v{\bf v}}^{\mathrm H})\right)(\mathbf{Q}^{\mathrm H}{\bf e}+{\bf{g}})+(1+\beta_{ r})\delta^{2}\triangleq{\Lambda}\label{eq1}.
\end{align}
\end{figure*}

\section{System Model}
\vspace{0.2cm}

Consider an  RIS-aided MISO downlink system,
 as shown in Fig. 1,
where the BS transmits signals to a single-antenna user.
The BS is equipped with $N$ antennas and the RIS has  $L$ reflecting elements.
Then, the signal transmitted from the BS is expressed as
\setcounter{equation}{0}
\begin{align}\label{0}
{\bf t} & ={\bf v}s+{\bf n}_{\mathrm t},
\end{align}
where $s\sim\mathcal{C}\mathcal{N}(0,1)$ is the data symbol and ${\bf v}\in{\mathbb{C}}^{N\times1}$ is the transmit beamforming vector.
In (\ref{0}), the hardware impairments at the BS is denoted as
${\bf {\bf n}}_{\mathrm t}\sim\mathcal{C}\mathcal{N}(0,\beta_{t}{\textrm{diag}}({\bf v{\bf v}}^{\mathrm H}))$,
 where $\beta_{ t}\in(0,1)$ is the proportionality coefficient which characterizes the level of the hardware impairments at the BS\cite{9239335}.

\begin{figure}
\includegraphics[scale=0.75]{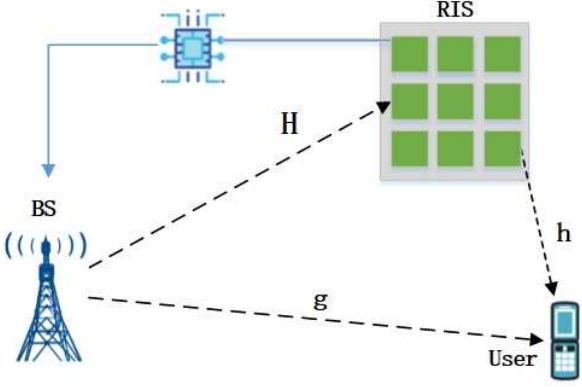}
\captionsetup{font={small},justification=raggedright}
\caption{ System model. }
\label{Figsysmodel}
\end{figure}


Let ${\bf{g}}\in\mathbb{C}^{N\times 1}$, $\mathbf{H}\in\mathbb{C}^{L\times N}$, and ${\bf h}\in{\mathbb{C}}^{L\times1}$
 represent the direct channel from the BS to the user, the channel from the BS to the RIS,  and  the channel  from the RIS to the user, respectively.
At the RIS, the reflection matrix is denoted as $\bm{\Phi}={\textrm {diag}}\left({e^{j\phi_{1}},e^{j\phi_{2}},\cdots,e^{j\phi_{L}}}\right)$,
where ${\phi}_i$ denotes the phase shift of the $i$-th reflecting element.
Therefore,
the received signal at the user can be written as
\begin{align}
y & =({\bf{g}}^{\mathrm H}+{\bf h}^{\mathrm H}\bm{\Phi}\mathbf{H})({\bf v}s+{\bf n}_{\mathrm t})+n+n_{\mathrm r}{}\nonumber\\
 &{} \triangleq \tilde{y}+n_{\mathrm r},
\end{align}
where $n\sim\mathcal{C}\mathcal{N}(0,\delta^{2})$ denotes the  additive white Gaussian noise (AWGN) received at the user.
The hardware impairments at the user is denoted as  ${n_{\mathrm r}} \sim \mathcal{C}\mathcal{N}(0,\beta_{r}\mathbb{E}\{|\tilde{y}|^{2}\}) $,
where $\beta_{ r}\in(0,1)$ is the proportionality coefficient characterizing the level of the hardware impairments at the user.
Let ${\bf e}=[{e^{j\phi_{1}},e^{j\phi_{2}},\cdots,e^{j\phi_{L}}}]^{\mathrm T}$ represent the reflecting beamforming vector containing the diagonal elements of $\bm{\Phi}$.
Define matrix ${\mathbf{Q}}=\textrm{diag}({\bf h}^{\mathrm H}){\mathbf{H}}$ as the cascaded channel matrix from the BS to the user through the RIS.
Then, we can obtain that ${\bf e}^{\mathrm H}\mathbf{Q}={\bf h}^{\mathrm H}\bm{\Phi}\mathbf{H}$.
Thus, the signal-to-noise ratio (SNR) $\gamma$ is expressed as
\begin{equation}\label{26}
\gamma=\frac{|({\bf{g}}^{\mathrm H}+{\bf e}^{\mathrm H}\mathbf{Q}){\bf v}|^{2}} {  {\mathbb E}\{ n_{\mathrm e} n_{\mathrm e}^*\}},
\end{equation}
where $n_{\mathrm e}=({\bf{g}}^{\mathrm H}+{\bf e}^{\mathrm H}\mathbf{Q}){\bf n}_{\mathrm t}+n+n_{\mathrm r}$ is the interference noise.
In \eqref{26}, ${\mathbb{E}\{n_{\mathrm e} n_{\mathrm e}^*\}}$ is the interference noise power,
which is given by (4) at the bottom of this page.

Due to the passive property of the RIS, it is  challenging to acquire the perfect CSI.
Hence, we consider the statistical CSI  error,
which means
that the CSI estimation error is random and follows certain distribution. Then, the cascaded channel and the direct
channel are respectively modeled as
\setcounter{equation}{4}
\begin{subequations} \label{7}
\begin{align}
\mathbf{Q}=\hat{\mathbf{Q}}+\Delta\mathbf{Q},\\
 \bf{g}=\hat{\bf{g}}+\Delta\bf{g},
\end{align}
\end{subequations}
where $\hat{\mathbf{Q}}$ and $\hat{\bf{g}}$ respectively represent the estimated cascaded channel and the estimated direct channel,
while $\Delta\mathbf{Q}$ and $\Delta\bf{g}$ are the corresponding channel estimation errors.
Based on this statistical CSI error model,
we assume that the channel estimation error vectors $\textrm{vec}(\Delta\mathbf{Q})$  and $\Delta\bf{g}$ follow the circularly
symmetric complex Gaussian (CSCG) distribution\cite{9130088}, and they are respectively given by
\begin{subequations} \label{1}
\begin{align}
&\textrm{vec}(\Delta\mathbf{Q})\sim{\cal C}{\cal N}({\bf 0},{\boldsymbol{{\mathbf{\Sigma}}}_{\rm q}}),\\
&\Delta{\bf{g}}\sim{\cal C}{\cal N}({\bf 0},{\boldsymbol{{\mathbf{\Sigma}}}}_{\rm g}),
\end{align}
\end{subequations}
where ${\boldsymbol{{\mathbf{\Sigma}}}_{\rm q}}\in\mathbb{C}^{LN\times LN}$ and ${\boldsymbol{{\mathbf{\Sigma}}}_{\rm g}}\in\mathbb{C}^{N\times N}$ are
the covariance matrices of the estimation error vectors.
\section{Problem Formulation and Robust Beamforming Designs }
\subsection{Problem Formulation}
In this work, the transmit power is minimized by jointly optimizing the transmit beamforming and the reflecting phase shifts,
subject to the unit-modulus constraints on the reflecting elements
and the outage probability constraint $\tau$, ${\tau}\in(0,1]$.
Thus, we can express the optimization problem as
\begin{subequations}\label{P2}
\begin{alignat}{2}
\min_{{\bf v},{\bf e}}\quad & \|{\bf v}\|_2^2\\
\mbox{s.t.}\quad & |e^{j\phi_{l}}|=1, \forall l=1,...,L,\label{17}\\\label{}
 & {\rm Pr}\left\{ \log_{2}(1+\gamma)\ge R\right\} \ge1-\tau,\label{16}
\end{alignat}
\end{subequations}
where $\|\cdot\|$ denotes the Euclidean norm.
The rate outage probability constraint (\ref{16}) guarantees
that the receiver can decode desired information at a data rate of $ R$ no less than the probability ${1-\tau}$.

Due to the fact that the constraints (\ref{17}) are non-convex and the constraint (\ref{16}) has no simple closed-form expression\cite{6891348},
it is challenging to solve  Problem (\ref{P2}). In order to address it,
we firstly perform mathematical transformations on constraint (\ref{16}).


Firstly, we can rewrite the rate outage probability as
\begin{align} \label{2}
 {\rm Pr}&\left\{ \log_{2}(1+\gamma)\ge R\right\}\nonumber \\
 & ={\rm Pr}\left\{ 1+\gamma \ge 2^{R} \right\} \nonumber\\
 & ={\rm Pr}\left\{\! ({\bf{g}}^{\mathrm H}\!+\! {\bf e}^{\mathrm H}{\mathbf{Q}}) {\bf v}{\bf {\bf v}}^{\mathrm H}({\mathbf{Q}}^{\mathrm H}{\bf e}\! +\!{\bf{g}})\ge (2^{R}-1){\Lambda} \right\}\nonumber \\
 & ={\rm Pr}\left\{ \!({\bf{g}}^{\mathrm H}\!+\!{\bf e}^{\mathrm H}{\mathbf{Q}})\mathbf{A}({\mathbf{Q}}^{\mathrm H}{\bf e}\!+\!{\bf{g}})\!-\!(1+\beta_{r})\delta^{2}\ge0\right\},
\end{align}
where $ \mathbf{A}=\left(\frac{1}{2^{R}-1}-\beta_{r}\right){\bf v}{\bf {\bf v}}^{\mathrm H}-(1+\beta_{r})\beta_{t}{\textrm{diag}}({\bf v{\bf v}}^{\mathrm H}) $.
By substituting (\ref{7}) into (\ref{2}),
the rate outage probability in (\ref{2}) is further written as
\begin{align}\label{3}
 {\rm Pr}&\left\{ ({\bf{g}}^{\mathrm H}+{\bf e}^{\mathrm H}{\mathbf{Q}})\mathbf{A}({\mathbf{Q}}^{\mathrm H}{\bf e}+{\bf{g}})-(1+\beta_{r})\delta^{2}\ge0\right\}\nonumber  \\
 & ={\rm Pr}\left\{ (\Delta{\bf{g}}^{\mathrm H}+{\bf e}^{\mathrm H}\Delta\mathbf{Q})\mathbf{A}(\Delta\mathbf{Q}^{\mathrm H}{\bf e}+\Delta{\bf{g}})\right.\nonumber\\
 &\left.+2{\rm Re}\left\{ (\hat{\bf{g}}^{\mathrm H}+{\bf e}^{\mathrm H}\hat{\mathbf{Q}})\mathbf{A}(\Delta\mathbf{Q}^{\mathrm H}{\bf e}+\Delta{\bf{g}})\right\}\right.\nonumber \\
 &\left. +(\hat{\bf{g}}^{\mathrm H}+{\bf e}^{\mathrm H}\hat{\mathbf{Q}})\mathbf{A}(\hat{\mathbf{Q}}^{\mathrm H}{\bf e}+\hat{\bf{g}})-(1+\beta_{r})\delta^{2}\ge0\right\}.
\end{align}
\vspace{-0.2cm}

By defining $\mathbf{E}={\bf e}{\bf e}^{\mathrm H}$, the first term in (\ref{3}) is rewritten as
\begin{align}
(&\Delta{\bf{g}}^{\mathrm H}+{\bf e}^{\mathrm H}\Delta\mathbf{Q})\mathbf{A}(\Delta\mathbf{Q}^{\mathrm H}{\bf e}+\Delta{\bf{g}})  \nonumber\\
&=\!\Delta\mathbf{g}^{\mathrm H}\!\mathbf{A}\Delta\mathbf{g}\!+\!2{\rm Re}\!\left\{ \!{\rm Tr}\{  \! {\bf e}^{\mathrm T}\!\Delta{\mathbf{Q}}^{*}\!\mathbf{A}^{\mathrm T}\!\Delta\mathbf{g} ^{*}\}\!\right\}\!\!+\! \!{\rm Tr}\{\!\mathbf{E}^{\mathrm T}\!\Delta{\mathbf{Q}}^{*}\!\!\mathbf{A}\!^{\mathrm T}\!\!\Delta\mathbf{Q}\!^{\mathrm T}\!\}    \nonumber \\
&\overset{(a)}{=}\!\Delta\mathbf{g}^{\mathrm H}\!\mathbf{A}\Delta\mathbf{g}+2{\rm Re}\left\{ \Delta{\mathbf{g}}^{\mathrm H}(\mathbf{A}\otimes{\bf e}^{\mathrm T})\mathrm{vec}(\Delta\mathbf{Q}^{*}) \right\}\nonumber\\
&\quad \quad \quad \quad +\mathrm{vec}^{\mathrm{T}}(\Delta\mathbf{Q})(\mathbf{A}\otimes\mathbf{E}^{\mathrm T})\mathrm{vec}(\Delta\mathbf{Q}^{*})  \nonumber \\
&\overset{(b)}{=}\zeta ^{2}_{\rm g}{\bf i}^{\mathrm H}_{\rm g}{\mathbf{A}}{\bf i}_{\rm g}\!+\!2{\rm Re}\left\{\zeta _{\rm g}\zeta _{\rm q}{\bf i}^{\mathrm H}_{\rm g}(\mathbf{A}\otimes{\bf e}^{\mathrm T}){\bf i}_{\rm q}^{*}  \right\}
\!+\!\zeta^{2}_{\rm q}{\bf i}_{\rm q}^{\mathrm T}(\mathbf{A}\otimes\mathbf{E}^{\mathrm T}){\bf i}_{\rm q}^{*}\nonumber \\
&={\bf i}^{\mathrm H}{\breve {\mathbf M}}{\bf i},\label{25}
\end{align}
where $(a)$ holds due to $\mathrm{Tr} \left\{ \mathbf{A}_{1}\mathbf{B}_{1}\mathbf{C}_{1}\mathbf{D}_{1}\right\}\!=\!(\mathrm{vec}(\mathbf{D}_{1}\!^{\mathrm T}))^{\mathrm T}(\mathbf{C}_{1}^{\mathrm T}\otimes\mathbf{A}_{1})\mathrm{vec}(\mathbf{B}_{1})$\cite{Zhang2017Matrix},
and $(b)$ holds by assuming the channel estimation error vectors ${\Delta\mathbf{g}}=\zeta_{\rm g} {\bf i}_{\rm g}\sim{\cal C}{\cal N}({\bf 0},\zeta_{\rm g}^{2}{\mathbf{I}_{N}})$
and $\mathrm{vec}(\Delta\mathbf{Q})=\zeta_{\rm q} {\bf i}_{\rm q}\sim{\cal C}{\cal N}({\bf 0},\zeta_{\rm q}^{2}{\mathbf{I}_{LN}})$,
where ${\bf i}_{\rm g}$ and ${\bf i}_{\rm q}$ are the CSCG random vectors,
 $\zeta_{\rm g}$ and $\zeta_{\rm q}$ are the constants which measure the relative amount of CSI uncertainties,
and ${{{\mathbf{I}}}_{N}}$ and ${{{\mathbf{I}}}_{LN}}$ are the $N\times N$ identity matrix and $LN\times LN$ identity matrix, respectively.
In \eqref{25}, ${\bf i}$ and ${\breve {\mathbf M}}$ are respectively defined as
\vspace{-0.2cm}
\begin{align}
{ {\bf i}}=[{\bf i}_{\rm g}^{\mathrm H} \quad {\bf i}_{\rm q}^{\mathrm T}]^{\mathrm H},
\begin{small}
{\breve {\mathbf M}}=
 \left[\!
 \begin{array}{ccc}
     {\zeta^{2}_{\rm g}}{\mathbf{A}} & \zeta_{\rm g} \zeta_{\rm q}(\mathbf{A}\otimes{\bf e}^{\mathrm T}) \\
     \zeta_{\rm g} \zeta_{\rm q}(\mathbf{A}\otimes{\bf e}^{*}) & \zeta^{2}_{\rm q}(\mathbf{A}\otimes{\mathbf E}^{\mathrm T})
 \end{array}
 \! \right].
\end{small}
\end{align}

\vspace{0.1cm}
Similarly, the second term of (\ref{3}) is rewritten as
\begin{align}
 &2{\rm Re}\left\{ (\hat{\bf{g}}^{\mathrm H}+{\bf e}^{\mathrm H}\hat{\mathbf{Q}})\mathbf{A}(\Delta\mathbf{Q}^{\mathrm H}{\bf e}+\Delta{\bf{g}})\right\} \nonumber\\
 &=2{\rm Re}\left\{\zeta_{\rm g}(\hat{\bf{g}}^{\mathrm H}+{\bf e}^{\mathrm H}\hat{\mathbf{Q}})\mathbf{A} {\bf i}_{\rm g}+ {\rm Tr}\{({\bf e} (\hat{\bf{g}}^{\mathrm H}\!+\!{\bf e}^{\mathrm H}\hat{\mathbf{Q}})\mathbf{A})^{\mathrm T}\!\Delta\mathbf{Q}^{*}\} \right\}\nonumber\\
 &\overset{(c)}{=}2{\rm Re}\left\{\zeta_{\rm g}(\hat{\bf{g}}^{\mathrm H}+{\bf e}^{\mathrm H}\hat{\mathbf{Q}})\mathbf{A} {\bf i}_{\rm g}+ \zeta_{\rm q}\mathrm{vec}^{\mathrm T}\big({\bf e}(\hat{\bf{g}}^{\mathrm H}+{\bf e}^{\mathrm H}\hat{\mathbf{Q}})\mathbf{A}\big){\bf i}_{\rm q}^{*}  \right \}\nonumber\\
 &=2{\rm Re}\left\{  {\breve {\bf m}}^{\mathrm H} {\bf i}\right \},
 \end{align}
where ${\breve {\bf m}}=
 \left[
 \begin{array}{ccc}
      \zeta_{\rm g} (\hat{\bf{g}}^{\mathrm H}\!+\!{\bf e}^{\mathrm H}\hat{\mathbf{Q}})\mathbf{A}   \quad \zeta_{\rm q}\mathrm{vec}^{\mathrm T}\big(\!{\bf e}(\hat{\bf{g}}^{\mathrm H}\!+\!{\bf e}^{\mathrm H}\hat{\mathbf{Q}})\mathbf{A}\!\big)
 \end{array}
  \right]^{\mathrm H}$, and $(c)$ is obtained by using ${\rm Tr}\{\mathbf{A}_{1}^{\mathrm T}\mathbf{B}_{1}  \} = \mathrm{vec}^{\mathrm T}(\mathbf{A}_{1})\mathrm{vec}(\mathbf{B}_{1})$\cite{Zhang2017Matrix}.
Then, 
the  constraint (\ref{16}) can be simplified as
\begin{equation}\label{9}
{\rm Pr}\left\{ {\bf i}^{\mathrm H}{\breve {\mathbf{M}}}{\bf i}+2\mathrm{Re}\left\{ {\breve {{\bf m}}}^{\mathrm H}{\bf i}\right\} +{\breve m}\ge0\right\} \ge1-\tau,
\end{equation}
where ${\breve m}=(\hat{\bf{g}}^{\mathrm H}+{\bf e}^{\mathrm H}\hat{\mathbf{Q}})\mathbf{A}(\hat{\mathbf{Q}}^{\mathrm H}{\bf e}+{\hat {\bf g}})-(1+\beta_{r})\delta^{2}$.
Then, we will further transform the constraint (\ref{16}) by utilizing Lemma 1 in \cite{6891348}, which is given by the following lemma.

\vspace{0.2cm}
\begin{lemma}
(Bernstein-Type Inequality: Lemma 1 in\cite{6891348})
 Assume $\mathit {\bf{a}}^{\mathrm H}{\mathbf M}{\bf{a}}+2{\rm Re}\left\{ {\bf m}^{\mathrm H}{\bf{a}}\right\}+m$, where ${\mathbf M}\in{\mathbb{H}}^{n\times n}, {\bf m}\in{\mathbb{C}}^{n\times 1},  m\in{\mathbb{R}}$ and ${\bf a}\in{\mathbb{C}}^{n\times 1}\sim{\cal C}{\cal N}({\bf 0},{\bf {I}})$. For any ${\tau}\in(0,1]$, $x$ and $y$ are slack variables,
then, we have the following relationship:
\begin{align}
 {\rm Pr}&\left\{ {\bf{a}}^{\mathrm H}{\mathbf M}{\bf{a}}+2{\rm Re}\left\{ {\bf m}^{\mathrm H}{\bf{a}}\right\}+m \ge 0 \right\}\ge 1-\tau{}\\
&\Rightarrow {}\begin{cases}
{\rm Tr}\left\{ \mathbf{M}\right\} -\sqrt{2\ln(1/\tau)}x-\ln(1/\tau)y+m\ge0 \\
\begin{gathered}
\begin{Vmatrix}\mathrm{vec}(\mathbf{M})\\
\sqrt{2}\bf{m}
\end{Vmatrix}_{2}\le x,\end{gathered}\\
y\mathbf{I}+\mathbf{M}\succeq\mathbf{0},y\ge0.\nonumber
\end{cases}
\end{align}
\end{lemma}

\vspace{-0.2cm}
By introducing the auxiliary variables $a$ and $b$,
we apply the Bernstein-Type Inequality in  Lemma 1
to transform the  original constraint (\ref{16}) as
\begin{small}
\begin{align}\label{18}
\begin{cases}
&{\rm Tr}\left\{{\breve {\mathbf{M}}}\right\} -\sqrt{2\ln(1/\tau)}a-\ln(1/\tau)b+{\breve m}\ge0, \\
&\begin{gathered}
\begin{Vmatrix}\mathrm{vec}({\breve {\mathbf{M}}})\\
\sqrt{2}{\breve {\bf{m}}}
\end{Vmatrix}_{2}\le x,\end{gathered}\\
&b\mathbf{I}+{\breve {\mathbf{M}}}\succeq\mathbf{0},b\ge0.
\end{cases}
\end{align}
\end{small}

The variables ${\rm Tr}\{\breve{ \mathbf{M}}\}$, $\|{\breve {\mathbf{M}}} \|_{F}$ and $\|{\breve{ \bf m}}\|^{2}_{2}$ in constraints (\ref{18}) can be further simplified as
\begin{small}
\begin{align}
 & \mathrm{Tr}\{\breve {\mathbf{M}} \} =  (\zeta ^{2}_{\rm g}+\zeta ^{2}_{\rm q}L){\rm Tr}\left\{ { \mathbf{A}} \right\} ,\nonumber \\
&\|\mathrm{vec}({\breve {\mathbf{M}}})\!\|_{2}\!=\!\|{\breve {\mathbf{M}}}\|_{F}\!=\!(\zeta ^{2}_{\rm g}\!+\!\zeta ^{2}_{\rm q}L)   \|\mathbf{A}\|_{F}\!=\!(\zeta ^{2}_{\rm g}\!+\!\zeta ^{2}_{\rm q}L)\|\mathrm{vec}({ {\mathbf{A}}})\|_{2},\nonumber\\
 &\|{\breve{\bf m}}\|^{2}_{2}= \zeta ^{2}_{\rm g}\|(\hat{\bf{g}}^{\mathrm H}+{\bf e}^{\mathrm H}\hat{\mathbf{Q}})\mathbf{A}\|^{2}_{2}\!+\! \zeta ^{2}_{\rm q}\!\|\mathrm{vec}^{\mathrm T}\!\big(\!{\bf e}(\hat{\bf{g}}^{\mathrm H}\!+\!{\bf e}^{\mathrm H}\hat{\mathbf{Q}}\!)\mathbf{A} \big)\|^{2}_{2} \nonumber\\
 &\quad\quad=\zeta ^{2}_{\rm g}\|(\hat{\bf{g}}^{\mathrm H}+{\bf e}^{\mathrm H}\hat{\mathbf{Q}})\mathbf{A}\|^{2}_{2}+\zeta ^{2}_{\rm q}\|{\bf e}(\hat{\bf{g}}^{\mathrm H}+{\bf e}^{\mathrm H}\hat{\mathbf{Q}})\mathbf{A}\|_{F}^{2}\nonumber\\
 &\quad\quad=\zeta ^{2}_{\rm g}\|(\hat{\bf{g}}^{\mathrm H}+{\bf e}^{\mathrm H}\hat{\mathbf{Q}})\mathbf{A}\|^{2}_{2}+\zeta ^{2}_{\rm q}L\|(\hat{\bf{g}}^{\mathrm H}+{\bf e}^{\mathrm H}\hat{\mathbf{Q}})\mathbf{A}\|_{2}^{2}\nonumber\\
 &\quad\quad=(\zeta ^{2}_{\rm g}+\zeta ^{2}_{\rm q}L)\|(\hat{\bf{g}}^{\mathrm H}+{\bf e}^{\mathrm H}\hat{\mathbf{Q}})\mathbf{A}\|^{2}_{2}.
\end{align}
\end{small}

\vspace{-0.5cm}
Then, Problem (\ref{P2}) can be approximated as
\vspace{-0.2cm}
\begin{subequations}\label{8}
\begin{align}
&\min_{{\bf v},{\bf e},a,b}\quad   \|{\bf v}\|_{2}^{2}&\\
&\quad\mbox{s.t.} \quad |e^{j\phi_{l}}|=1, \forall l=1,...,L,&\\
&\quad \quad \quad (\zeta_{\rm g}^{2}+\zeta_{\rm q} ^{2}L){\rm Tr}\left\{ \mathbf{A}\right\}-\sqrt{2\ln(1/\tau)}a \nonumber\\
&\quad \quad \quad \quad \quad \quad-\ln(1/\tau)b+{\breve m}\ge0,\label{24}\\
 & \quad\quad\quad\begin{gathered}\begin{Vmatrix}(\zeta _{\rm g}^{2}+\zeta _{\rm q}^{2}L)\mathrm{vec}(\mathbf{A})\\
\sqrt{2(\zeta _{\rm g}^{2}+\zeta _{\rm q}^{2}L)} \mathbf{A}(\hat{\bf{g}}+\hat{\mathbf{Q}}^{\mathrm H}{\bf e})
\end{Vmatrix}_{2}\le a,\end{gathered}\label{10}
\\
 & \quad\quad\quad b\mathbf{I}+(\zeta _{\rm g}^{2}+\zeta _{\rm q}^{2}L)\mathbf{A}\succeq\mathbf{0},b\ge0. \label{23}
\end{align}
\end{subequations}

\subsection{Optimizing the Transmit Beamforming Vector}
Due to the fact that the vector ${\bf v}$ and the vector ${\bf e}$ are coupled together,
Problem (\ref{8}) is non-convex.
 To solve it, we use the alternating optimization (AO) technique to optimize the  vector ${\bf v}$ and the  vector ${\bf e}$.
 Firstly, we optimize
${\bf v}$ when ${\bf e}$ is fixed.
By defining $\mathbf{V}={\bf v}{\bf v}^{\mathrm H}$, Problem (\ref{8}) is reformulated as
\vspace{-0.2cm}
\begin{subequations}\label{13}
\begin{align}\min_{\mathbf{V},a,b}\quad & {\rm Tr}\left\{ \mathbf{V}\right\} \\
\mbox{s.t.}\quad & \eqref{24}, \eqref{10}, \eqref{23},\\
 & \mathbf{V}\succeq\mathbf{0},\mathrm{rank}(\mathbf{V})=1.
\end{align}
\end{subequations}

\vspace{-0.1cm}
By utilizing the semidefinite relaxation (SDR) method\cite{9180053}, the constraint $\mathrm{rank}(\mathbf{V})=1$ is removed.
Therefore, the resulting  semidefinite programming (SDP)
problem can be effectively solved  by  utilizing the CVX tools\cite{cvx}.
Finally, the optimal  $\bf v$ can be obtained from $\mathbf{V}$ by using
Gaussian randomization method\cite{9180053}.

\subsection{Optimizing the Phase Shifts Vector }
 In this subsection, we will optimize the phase shifts vector ${\bf e}$ when the transmit beamforming vector ${\bf v}$ is fixed.
 Due to the fact that the objective function $\|{\bf v}\|_{2}^{2}$ of Problem (\ref{8}) does not contain the phase shifts vector ${\bf e}$,
Problem (\ref{8}) reduces to a feasibility check problem when the variables $\{{\bf v}, a, b\}$ are fixed.
By adopting the similar method in\cite{9180053}, we introduce a slack variable $\mu \ge 0$,
and it can achieve a strictly larger rate than the target rate $R$ for the user
and accelerate the convergence.
Hence, the rate outage probability in constraint (\ref{2}) is modified as
\vspace{-0.2cm}
\begin{align}\label{21}
\begin{small}
{\rm Pr}\left\{ ({\bf{g}}^{\mathrm H}+{\bf e}^{\mathrm H}{\mathbf{Q}})\mathbf{A}({\mathbf{Q}}^{\mathrm H}{\bf e}+{\bf{g}})\!-\!(1+\beta_{r})\delta^{2}\!\!-\!\mu\ge0\right\}.
\end{small}
\end{align}

Then, by performing the same transformations for (\ref{21}) again, we can obtain the approximation of (\ref{21}) as 
\vspace{-0.2cm}
\begin{small}
\begin{equation}
\begin{split}
\begin{cases}
&\!{\rm Tr}\left\{\breve {\mathbf{M}}\right\}\!-\!\sqrt{2\ln(1/\rho)}a\!-\!\ln(1/\rho)b\!+\!{\breve m}_{{\bf e}}\ge0,\\
&\eqref{10}, \quad \eqref{23},
\end{cases}
\end{split}
\end{equation}
\end{small}
\hspace{-0.15cm}where ${\breve m}_{{\bf e}}=(\hat{\bf{g}}^{\mathrm H}+{\bf e}^{\mathrm H}\hat{\mathbf{Q}})\mathbf{A}(\hat{\mathbf{Q}}^{\mathrm H}{\bf e}+{\hat {\bf g}})-(1+\beta_{r})\delta^{2}-\mu$.
By defining ${\tilde{{\mathbf{E}}}}=
 \left[
  \begin{array}{ccc}
     {\bf{e}}{\bf{e}}^{\mathrm H}    & {\bf{e}}  \\
     {\bf{e}}^{\mathrm H} & 1 \nonumber
 \end{array}
 \right]$, we can rewrite  ${\breve m}_{{\bf e}}$ as $
{\breve m}_{{\bf e}}={\rm Tr}\left\{\!\mathbf{B}{\tilde{\mathbf{E}}}\right\}\!+\!\hat{\bf{g}}^{\mathrm H}\mathbf{A}\hat{\bf{g}}\!-\!(1\!+\!\beta_{r})\delta^{2}\!-\!\mu,
$
where
$
 {\mathbf B}=
 \left[
 \begin{array}{ccc}
     \hat{\mathbf{Q}}\mathbf{A}\hat{\mathbf{Q}}^{\mathrm H} & \hat{\mathbf{Q}}\mathbf{A}\hat{\bf{g}}  \\
     \hat{\bf{g}}^{\mathrm H}\mathbf{A}\hat{\mathbf{Q}}^{\mathrm H} & 0 \nonumber
 \end{array}
  \right].
$
\begin{algorithm}[t] 
\caption{{\textcolor{black}{AO algorithm for Problem (\ref{8})}}} 
\begin{algorithmic}[1] 
\Require  \mbox{Initial iteration number $n=0$, ${\bf e}^{(0)}$ and ${\bf v}^{(0)}$.}
\Repeat
\State Given ${\bf e}^{(n)}$, calculate  ${\bf v}^{(n+1)}$ from Problem (\ref{13});
\State Given ${\bf v}^{(n+1)}$, calculate  ${\bf e}^{(n+1)}$ from Problem (\ref{12});
\State Set $n \leftarrow n+1$;
\Until{The value $\|{\bf v}\|^2_2 $ converges.}
\end{algorithmic}
\end{algorithm}
In order to obtain the variable ${\tilde {\mathbf{E}}}$,
 the constrsint (\ref{10}) is  reformulated as
\begin{equation}\label{11}
\begin{small}
 \!(\zeta _{\rm g}^{2}\!\!+\!\zeta _{\rm q}^{2}L)^{2}\!\|\mathbf{A}\|_{F}^{2}\!+\!2(\zeta _{\rm g}^{2}\!+\!\zeta _{\rm q}^{2}L)\!({\rm Tr}\left\{{ \mathbf{C}}{\tilde{\mathbf{E}}}\right\}\!+\!\hat{\bf g}\!^{\mathrm{H}}\!{\mathbf{A}}\!{\mathbf{A}}\!^{\mathrm{H}}\!\hat{\bf g})\! 
\le \!a^{2}\!,
\end{small}
\end{equation}
where
$\small{
 {\mathbf C}=
 \left[
 \begin{array}{ccc}
     \hat{\mathbf{Q}}\mathbf{A}\mathbf{A}^{\mathrm H}\hat{\mathbf{Q}}^{\mathrm H} & \hat{\mathbf{Q}}\mathbf{A}\mathbf{A}^{\mathrm H}\hat{\bf{g}}  \\
     \hat{\bf{g}}^{\mathrm H}\mathbf{A}\mathbf{A}^{\mathrm H}\hat{\mathbf{Q}}^{\mathrm H} & 0
 \end{array}
  \right].\nonumber}
$

Since constraint (\ref{11}) is non-convex, we need to linearize it  by using the first-order linear approximation.
Then, the constraint (\ref{11}) is reformulated as
\begin{align}\label{6}
 &
 (\zeta _{\rm g}^{2}+\zeta _{\rm q}^{2}L)^{2}\|\mathbf{A}\|_{F}^{2}+2(\zeta _{\rm g}^{2}+\zeta _{\rm q}^{2}L)({\rm Tr}\left\{{ \mathbf{C}}{\tilde{\mathbf{E}}}\right\}+\hat{\bf  g}^{\mathrm{H}}{\mathbf{A}}{\mathbf{A}}^{\mathrm{H}}\hat{\bf g}) \nonumber \\
&\quad \quad \quad \quad \quad \quad \quad \quad \quad \quad \quad \quad\le2a^{(n)}a-|a^{(n)}|^{2},
\end{align}
where $a^{(n)}$ on the right hand side of the above inequality is the optimal solution in the $n$-th iteration.
By optimizing  $\tilde {\mathbf E}$ to make the achievable rate  larger than the target rate $R$,
Problem (\ref{13}) has more space to reduce the transmit power.
Therefore, the phase shifts optimization problem
is reformulated as
\vspace{-0.2cm}
\begin{small}
\begin{subequations}\label{12}
\begin{align}&\max_{{\tilde{\mathbf E}},\mu,a,b}\quad \mu&\\
\vspace{-0.3cm}
&\quad\mbox{s.t.}\quad  (\zeta _{\rm g}^{2}+\zeta _{\rm q}^{2}L){\rm Tr}\left\{ \mathbf{A}\right\} -\sqrt{2\ln(1/\rho)}a\nonumber\\
&\quad\quad\quad\quad\quad-\ln(1/\rho)b+{\breve m}_{{\bf e}}\ge0,\\
&\quad\quad\quad(\ref{6}), \mu\ge0, a\ge0,\\
&\quad\quad\quad{\tilde {\mathbf{E}}}\succeq\mathbf{0}, \!\mathrm{rank}({\tilde {\mathbf{E}}})\!=\!1\!, \![{\tilde {\mathbf{E}}}]_{l,l}=1,l=1,...,L+1.
\end{align}
\end{subequations}
\end{small}
\hspace{-0.2cm}where $[\tilde {\mathbf {E}}]_{i,j}$ denotes  the $(i, j)$th element of $\tilde {\mathbf E}$.
To solve Problem (\ref{12}), the SDR  is employed.
Then, Problem (\ref{12}) becomes
a convex SDP problem.
By making use of the CVX tools, we can obtain the optimal solution $\tilde {\mathbf{E}}$ to Problem (\ref{12}).
The optimal phase shifts vector $\bf{e}$  can  be also obtained
from the optimal $\tilde {\mathbf{E}}$  by using Gaussian randomization techniques.
Algorithm 1 presents the overall algorithm of AO method for solving Problem (\ref{8}).
By iterating step 2 and step 3 alternately,
the transmit power decreases monotonically until the convergence accuracy is reached.
 By adopting the similar method in \cite{9180053},
 we can obtain that
 the approximate complexity of Problem (\ref{13}) is ${\cal O_{\mathbf{V}}}\left( (2N+2)^{1/2}n_{1}[n_{1}^2+2n_{1}{N}^2+2{N}^3+n_{1}{N}^{2}(N+1)^{2}] \right)$,
 where $n_{1}=N^2$.
Similarly,
the complexity of Problem (\ref{12}) is ${\cal O_{\tilde {\mathbf{E}}}}\left( (4+2L)^{1/2}n_{2}[n_{2}^2+n_{2}({L}^2+ {N}^{2}(N+1)^{2}+L)] \right)$, where $n_{2}=L^2$.
 Finally, the approximate complexity of the proposed algorithm is ${\cal O_{\mathbf{V}}}+{\cal O_{\tilde {\mathbf{E}}}}$.
 \vspace{0.3cm}

\begin{figure}
\vspace{-0.6cm}
\centering
\includegraphics[scale=0.34]{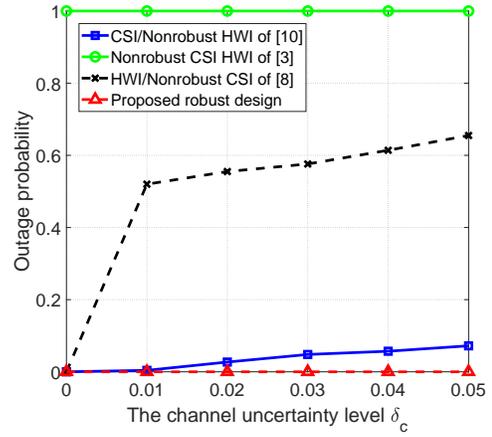}
\captionsetup[figure]{name={Fig.},labelsep=period}
\captionsetup{justification=raggedright}
\vspace{-0.4cm}\caption{ Outage probability versus the channel uncertainty level $\delta_c$ when $N=2$ and $M=24$.}
\end{figure}
\vspace{-0.25cm}

\section{Simulation Results}

In this section, we provide numerical results to evaluate the performance of our proposed algorithm.
We assume that the BS, the RIS, and the user are respectively located at (0 m, 0 m), (90 m, 0 m), and (90 m, 5 m) in a two-dimensional plane.
 The large-scale fading of the channels are
modeled as $\small {{\rm{PL}}\! =\!-30\!  - \!10\alpha {\log _{10}}d}$, where $\alpha$ is the path loss exponent and $d$ is the link distance in meter.
In this work, we set $\alpha = 3$ and $\alpha = 4$
for the cascaded  channel and  the direct channel, respectively.
The small-scale fading  is assumed to be Rician distributed. 
For simplicity, the Rician factor is assumed to be 5.
We respectively define  
$\zeta^{2}_{\rm q}=\delta^{2}_c||\textrm{vec}(\hat{\mathbf{Q}})||^{2}_{2}$ and $\zeta^{2}_{\rm g}=\delta^{2}_c||\hat{\bf{g}}||^{2}_{2}$,
 where $\delta_c \in[0,1)$  is the
channel uncertainty level.
We assume that the BS and the user have the same level of the hardware impairments, i.e., $\beta_{ t}=\beta_{ r}=\beta$.
The other parameters are set as: noise power of $\delta^{2}=-80$ dBm,  target rate of $R=1.5$ bit/s/Hz, convergence tolerance of $\epsilon =10^{-6}$, outage probability of $\tau=0.01$.



\begin{figure}
\centering
\vspace{-1.5cm}
\includegraphics[scale=0.33]{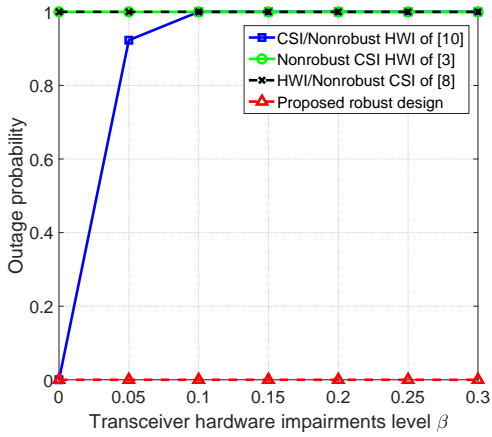}
\captionsetup{justification=raggedright}
\vspace{-0.6cm}
\caption{Outage probability versus the transceiver hardware impairments level $\beta$ when $N=2$ and $M=24$. }
\end{figure}


We illustrate the advantage of our proposed robust beamforming design by comparing it with the following  schemes:

$\bullet$
HWI/Nonrobust CSI of \cite{9239335}:
It only considered the impact of the transceiver hardware impairments in the transmission beamforming design,
while the  channel estimation error was ignored.
which corresponds  to the beamforming design scheme in \cite{9239335}.

$\bullet$ CSI/Nonrobust HWI of \cite{9180053}:
The channel estimation error was only considered in the robust transmission design,
while the  hardware impairments were ignored,
which corresponds to the beamforming design scheme in \cite{9180053}.




$\bullet$ Nonrobust CSI HWI of \cite{8811733}:
Both channel estimation error and hardware impairments were ignored during the transmission design,
which corresponds to the beamforming design scheme in \cite{8811733}.

Fig. 2 illustrates the outage probability versus the  CSI uncertainty level.
As seen from Fig. 2,
the outage probabilities of both ``CSI/Nonrobust HWI of \cite{9180053}'' and ``HWI/Nonrobust CSI of \cite{9239335}'' gradually increase with the increase of
 channel estimation error.
In particular, the outage probabilities of 
 ``Nonrobust CSI HWI of \cite{8811733}'' are always 1, which is due to the fact that the transceiver hardware impairments always deteriorate the
 the received signal quality.
By contrast, our proposed robust design method can always ensure that the outage probability is 0 with both transceiver hardware
impairments and imperfect CSI. Fig. 3 depicts the outage probability versus the transceiver hardware impairments level.
We can also find that our proposed robust design can keep the outage probability at 0.
The system performances for the other three schemes become worse with the increase of transceiver hardware impairments level.


Fig. 4 depicts the transmit power versus the number of RIS reflecting elements $L$.
Compared our proposed robust design
with other three schemes and ``Perfect CSI HWI'', which means perfect CSI acquisition and ideal hardware,
it is observed that the transmit power of our proposed design
is higher than others.
Since the robust transmission design takes into account both the hardware impairments and the imperfect CSI,
the BS needs higher power to transmit data in order to compensate  for  the  signal loss.


\begin{figure}
\vspace{-1.5cm}
\centering
\includegraphics[scale=0.33]{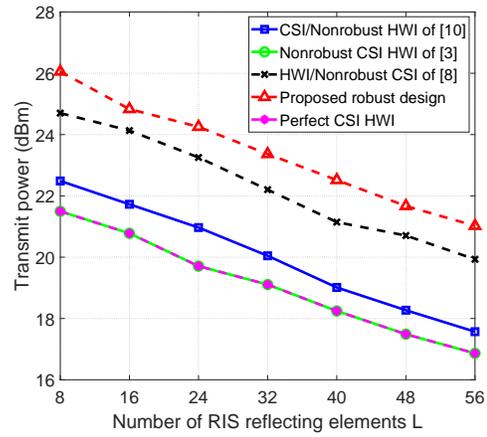}
\captionsetup{justification=raggedright}
\vspace{-0.6cm}
\caption{ Transmit power versus the number of RIS reflecting elements.}
\end{figure}

\vspace{-0.2cm}
\section{Conclusions}
This work studied an RIS-aided wireless  system,
where the impacts of both transceiver hardware impairments and imperfect CSI were considered.
Specifically,
the transmit power of the BS is minimized,
subject to the outage probability constraint and the unit-modulus
constraints on the reflecting elements.
By adopting the Bernstein-Type Inequality, we reformulated the constraints into the tractable forms.
Then, the reformulated problem was solved via the AO framework.
Simulation results demonstrated that the robustness of the  proposed  transmission  design  with both hardware impairments and imperfect CSI.

\bibliographystyle{IEEEtran}
\bibliography{IEEEabrv,myref}

\end{document}